# Adaptive MIMO Channel Estimation using Sparse Variable Step-Size NLMS Algorithms

Guan Gui[1], Li Xu[1], Lin Shan[2], and Fumiyuki Adachi[3]

1. Department of Electronics and Information Systems, Akita Prefectural University, Akita, Japan
2. Wireless Network Research Institute, National Institute of Information and Communication Technology, Yokosuka, Japan
3. Department of Communications Engineering, Graduate School of Engineering, Tohoku University, Sendai, Japan
Emails: {guiguan,xuli}@akita-pu.ac.jp, shanlin@nict.go.jp, adachi@ecei.tohoku.ac.jp

*Abstract*—To estimate multiple-input multiple-output (MIMO) channels, invariable step-size normalized least mean square (ISSNLMS) algorithm was applied to adaptive channel estimation (ACE). Since the MIMO channel is often described by sparse channel model due to broadband signal transmission, such sparsity can be exploited by adaptive sparse channel estimation (ASCE) methods using sparse ISS-NLMS algorithms. It is well known that step-size is a critical parameter which controls three aspects: algorithm stability, estimation performance and computational cost. The previous approaches can exploit channel sparsity but their step-sizes are keeping invariant which unable balances well the three aspects and easily cause either estimation performance loss or instability. In this paper, we propose two stable sparse variable step-size NLMS (VSS-NLMS) algorithms to improve the accuracy of MIMO channel estimators. First, ASCE for estimating MIMO channels is formulated in MIMO systems. Second, different sparse penalties are introduced to VSS-NLMS algorithm for ASCE. In addition, difference between sparse ISSNLMS algorithms and sparse VSS-NLMS ones are explained. At last, to verify the effectiveness of the proposed algorithms for ASCE, several selected simulation results are shown to prove that the proposed sparse VSS-NLMS algorithms can achieve better estimation performance than the conventional methods via mean square error (MSE) and bit error rate (BER) metrics.

## I. INTRODUCTION

High-rate data broadband transmission over multiple-input multiple-output (MIMO) channel is becoming one of mainstream techniques for the next generation communication systems [1], [2]. The major motivation is due to the fact that MIMO technology is a way of using multiple antennas to simultaneously transmit multiple streams of data in wireless communications [3] and hence it can bring considerable improvements such as data rate, reliability and energy efficiency. However, coherent receivers require accurate channel state information (CSI) since the received signals are distorted by multipath fading transmission. The accurate estimation of channel impulse response (CIR) is a crucial aspect and challenging issue in coherent modulation and its accuracy has a significant impact on the overall performance of the communication system.

During last decades, based on the assumption of dense CIRs, linear channel estimation methods, e.g., least squares (LS), were proposed for MIMO systems. By applying these approaches, the performance of linear methods depend only on the size of MIMO channel. Note that narrowband MIMO channel may be modeled as dense channel model because of its very short time delay spread; however, broadband MIMO channel is often modeled as sparse channel model [4]. A typical example of sparse channel is shown in Fig. 1.

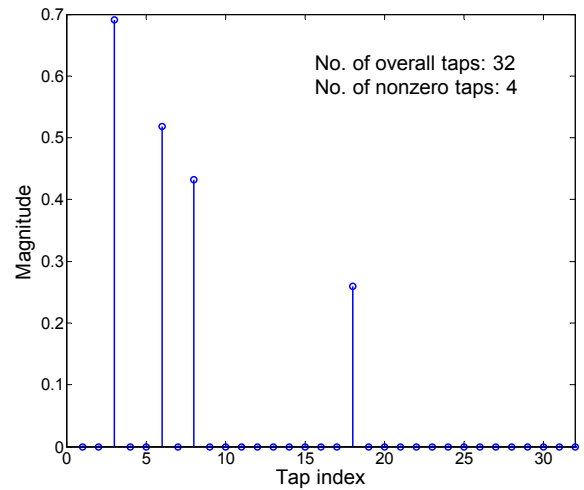

Fig. 1. A typical example of sparse multipath channel.

Adaptive sparse channel estimation (ASCE) methods using sparse invariable step-size (ISS) least mean square algorithms (ISS-LMS) were proposed in [5]–[7] for single-input single output (SISO) channels. However, conventional ISS-LMS methods have two main drawbacks: 1) sensitive to random scale of training signal and 2) unstable in low signal-to-noise ratio (SNR) regime. To overcome the two harmful factors on channel estimation and extend their applications to estimate MIMO channels, sparse ISS normalized least mean square (ISS-NLMS) algorithms, e.g., zero-attracting ISS-NLMS (ZA-ISS-NLMS) and reweight ZA-ISS-NLMS (RZA-ISS-NLMS), were proposed in [8]. It is well known that step-size is a critical parameter which controls the estimation performance, convergence rate and computational cost. Different from conventional sparse ISS-NLMS algorithms [5]–[8], zero-attracting variable step size NLMS (ZA-VSS-NLMS) algorithm was proposed for ASCE to improve estimation performance in sparse multipath single-input single-output (SISO) systems [9]. Unlike the previous works, this paper proposes two sparse VSS-NLMS algorithms for estimating sparse MIMO channels. The main contribution of this paper is summarized as follows. First, we extend the method in [9] from SISO to MIMO systems.

Second, a re-weighted ZA-VSS-NLMS (RZA-VSS-NLMS) is proposed to further improve the estimation performance of MIMO channels. In addition, we explain the reason why sparse VSS-NLMS algorithms can achieve better performance than conventional sparse ISS-NLMS ones. Finally, Monte Carlo based computer simulations are conducted to confirm the effectiveness of our proposed algorithms via two metrics: bit error rate (BER) and mean square error (MSE).

The remainder of this paper is organized as follows. A baseband MIMO system model is described and problem formulation is presented in Section II. In section III, sparse ISS-NLMS algorithms are reviewed and sparse VSS-NLMS algorithms are proposed. A figure example is also given to explain the difference between ISS and VSS based algorithms. Simulation results are presented in section IV in order to assess the proposed methods. Finally, we conclude the paper in Section V.

*Notation*: Throughout the paper, capital bold letters and small bold letters denote matrices and row/column vectors, respectively; The discrete Fourier transform (DFT) matrix is denoted by $\boldsymbol{F}$ with entries $[\boldsymbol{F}]_{kq} = 1/K \, e^{-j2\pi kq/K}$, $k, q = 0, 1, ..., K-1$; Matrices and vectors are represented by boldface upper case letters and boldface lower case letters, respectively; The superscripts $(\cdot)^T$, $(\cdot)^H$, $Tr(\cdot)$ and $(\cdot)^{-1}$ denote the transpose, the Hermitian transpose, the trace and the inverse operators, respectively; $E(\cdot)$ denotes the expectation operator; $\|\boldsymbol{h}\|_0$ is the $\ell_0$ norm operator that counts the number of nonzero taps in $\boldsymbol{h}$ and $\|\boldsymbol{h}\|_p$ stands for the $\ell_p$ norm operator which is computed as $\|\boldsymbol{h}\|_p = (\sum_i |h_i|^p)^{1/p}$, where $p \in \{1, 2\}$ is considered in this paper; $\text{sgn}(\cdot)$ is a component-wise function which is defined by $\text{sgn}(h) = 1$ for $h > 0$, $\text{sgn}(h) = 0$ for $h = 0$, and $\text{sgn}(h) = -1$ for $h < 0$.

## II. SYSTEM MODEL

A frequency-selective fading MIMO communication system using OFDM modulation scheme is considered. Initially, frequency domain training signal vector $\overline{\boldsymbol{x}}_{n_t}(t) = [\overline{x}_{n_t}(t, 0), ..., \overline{x}_{n_t}(t, L-1)]^T$, $n_t = 1, 2, ..., N_t$ is fed to inverse DFT (IDFT) at the $n_t$-th antenna, where $L$ is the number of subcarriers and $N_t$ is the number of transmit antenna. Assume that the transmit power is normalized as $E\|\overline{\boldsymbol{x}}_{n_t}\|_2^2 = 1$. The resultant vector $\boldsymbol{x}_{n_t}(t) = \boldsymbol{F}^H \overline{\boldsymbol{x}}_{n_t}(t)$ is padded with cyclic prefix (CP) of length $L_{CP} \geq L$ to avoid inter-block interference (IBI). After CP removal, the received signal vector at the $n_r$-th antenna for time $t$ is written as $y_{n_r}(t)$, where $n_r = 1, 2, ..., N_r$. Then, the received signal vector $\boldsymbol{y}$ and input signal vector $\boldsymbol{x}(t)$ are related by

$$y_{n_r}(t) = \sum_{n_t=1}^{N_t} \boldsymbol{h}_{n_r n_t}^T \boldsymbol{x}_{n_t}(t) + z_{n_r}(t) \\ = \boldsymbol{h}_{n_r:}^T \boldsymbol{x}(t) + z_{n_r}(t), \quad (1)$$

where $\boldsymbol{x}(t) = [\boldsymbol{x}_1^T(t), \boldsymbol{x}_2^T(t), ..., \boldsymbol{x}_{N_t}^T(t)]^T$ collects all of the input signal vectors from different antennas at the transmitter;

$z_{n_r}(t)$ is an additive white Gaussian noise (AWGN) variable with distribution $\mathcal{CN}(0, \sigma_n^2)$ and $n_r$-th received multiple-input single-output (MISO) channel vector $\boldsymbol{h}_{n_r:}$ is written as

$$\boldsymbol{h}_{n_r:} = [\underbrace{h_{n_r 1, 0}, \cdots, h_{n_r 1, L-1}}_{\boldsymbol{h}_{n_r 1}^T}, \cdots, \underbrace{h_{n_r n_t, 0}, \cdots, h_{n_r n_t, L-1}}_{\boldsymbol{h}_{n_r n_t}^T}, \\ \cdots, \underbrace{h_{n_r N_t, 0}, \cdots, h_{n_r N_t, L-1}}_{\boldsymbol{h}_{n_r N_t}^T}]^T, \quad (2)$$

and the matrix-vector form of system model (1) is also written as

$$\boldsymbol{y}(t) = \boldsymbol{H}\boldsymbol{x}(t) + \boldsymbol{z}(t), \quad (3)$$

where received signal vector $\boldsymbol{y}(t)$, noise vector $\boldsymbol{z}(t)$ and channel matrix $\boldsymbol{H}$ can be represented, respectively, as follows:

$$\boldsymbol{y}(t) = [y_1(t), y_2(t), \cdots, y_{N_r}(t)]^T, \quad (4)$$

$$\boldsymbol{z}(t) = [z_1(t), z_2(t), \cdots, z_{N_r}(t)]^T, \quad (5)$$

$$\boldsymbol{H} = \begin{bmatrix} \boldsymbol{h}_{1:}^T \\ \boldsymbol{h}_{2:}^T \\ \vdots \\ \boldsymbol{h}_{N_r:}^T \end{bmatrix} = \begin{bmatrix} \boldsymbol{h}_{11}^T & \boldsymbol{h}_{12}^T & \cdots & \boldsymbol{h}_{1N_t}^T \\ \boldsymbol{h}_{21}^T & \boldsymbol{h}_{22}^T & \cdots & \boldsymbol{h}_{2N_t}^T \\ \vdots & \vdots & \ddots & \vdots \\ \boldsymbol{h}_{N_r 1}^T & \boldsymbol{h}_{N_r 2}^T & \cdots & \boldsymbol{h}_{N_r N_t}^T \end{bmatrix} \quad (6)$$

where $\boldsymbol{h}_{n_r n_t}$, $n_r = 1, 2, \cdots, N_t$, is assumed equal $L$-length sparse channel vector from receiver to $n_t$-th antenna. In addition, we also assume that each channel vector $\boldsymbol{h}_{n_r n_t}$ is only supported by $T$ dominant channel taps.

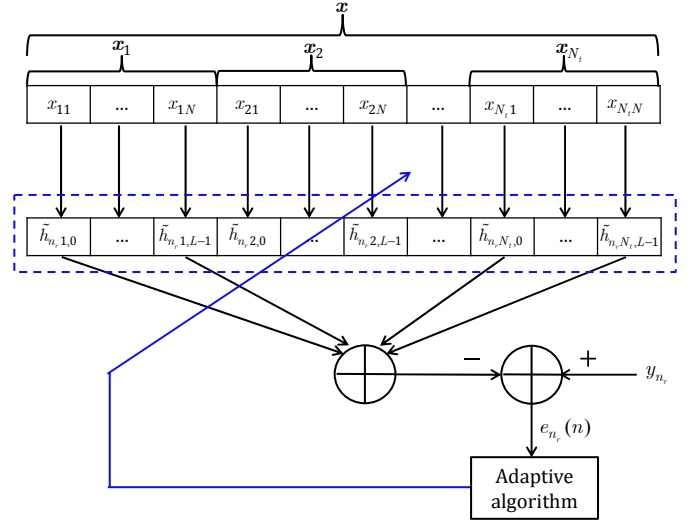

Fig. 2. MISO channel estimation at $n_r$-th antenna of the receiver.

## III. ADAPTIVE CHANNEL ESTIMATION MEHTODS

According to the system model in Eq. (1), the $n$-th updating estimation error $e_{n_r}(n)$ can be written as

$$e_{n_r}(n) = y_{n_r}(t) - \tilde{y}_{n_r}(n) = y_{n_r}(t) - \tilde{\boldsymbol{h}}_{n_r:}^T(n)\boldsymbol{x}(t), \quad (7)$$

where $\tilde{\boldsymbol{h}}_{n_r:}(n)$ denotes an MISO channel estimator of the $\boldsymbol{h}_{n_r:}(n)$; $\boldsymbol{e}(n) = [e_1(n), e_2(n), \cdots, e_{N_r}(n)]^T$ denotes receive error vector at the $n$-th adaptive update; and $y_{n_r}(t)$ is the receive signal at the $n_r$-th receive antenna.

*A. Review of ZA-ISS-NLMS and RZA-ISS-NLMS*

For estimating $h_{n_r:}(n)$ as shown in Fig. 3, ZA-ISS-NLMS [7] filtering algorithm was proposed as

$$\tilde{h}_{n_r:}(n+1) = \tilde{h}_{n_r:}(n) + \frac{\mu e_{n_r}(t)x(t)}{x^T(t)x(t)} - \gamma_{ZA}\,\mathrm{sgn}\!\left(\tilde{h}_{n_r:}(n)\right) \quad (8)$$

where $\gamma_{ZA} = \mu\lambda_{ZA}$, $\lambda_{ZA} > 0$ is a regularization parameter to balance the square estimation error $e_{n_r}(t)$ and sparse penalty of $\tilde{h}_{n_r:}(n)$. Motivated by reweighted $\ell_1$-norm minimization recovery algorithm [10], Chen et al. have proposed a heuristic approach to reinforce the zero attractor which was termed as the RZA-ISS-LMS [11]. RZA-ISS-NLMS [7] was proposed as

$$\tilde{h}_{n_r:}(n+1) = \tilde{h}_{n_r:}(n) + \frac{\mu e_{n_r}(t)x(t)}{x^T(t)x(t)} - \frac{\gamma_{RZA}\,\mathrm{sgn}\!\left(\tilde{h}_{n_r:}(n)\right)}{1+\varepsilon_{RZA}\left|\tilde{h}_{n_r:}(n)\right|}, \quad (9)$$

where $\gamma_{RZA} = \mu\lambda_{RZA}\varepsilon_{RZA}$, $\lambda_{RZA} > 0$ is the regularization parameter and reweighted factor $\varepsilon_{RZA} > 0$ is the positive threshold. Recall that the ZA-ISS-NLMS algorithm in Eq. (8) does not make use of the VSS rather than ISS. Inspired by the VSS-NLMS algorithm which has been proposed in [12], to improve estimation performance of MIMO channels, sparse VSS-NLMS algorithms are proposed. Unlike the sparse ISS-NLMS algorithm, sparse VSS-NLMS algorithms are time-variant with respect to the accuracy of updating estimators.

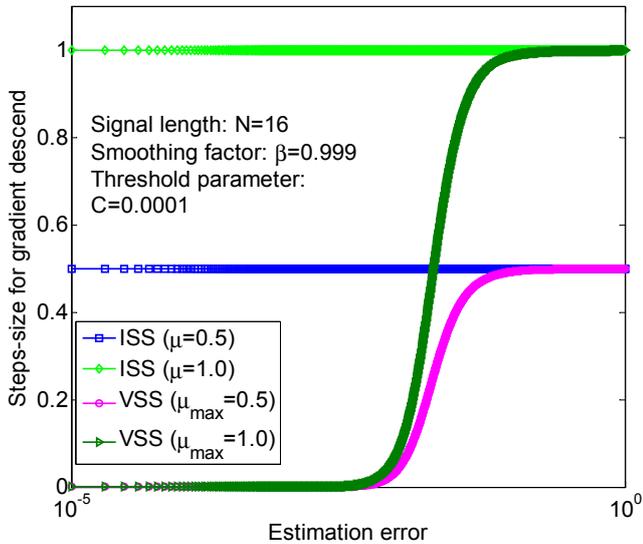
Fig. 3. ISS and VSS versus updating estimation error.

*B. ZA-VSS-NLMS*

At time $t$, based on the previous research on the ZA-ISS-NLMS and VSS-NLMS algorithms, ZA-VSS-NLMS algorithm is proposed as follows:

$$\tilde{h}_{n_r:}(n+1) = \tilde{h}_{n_r:}(n) + \frac{\mu_{n_r}(n)e_{n_r}(n)x(t)}{x^T(t)x(t)} - \gamma_{ZA}\,\mathrm{sgn}\!\left(\tilde{h}_{n_r:}(n)\right), \quad (10)$$

where $\mu_{n_r}(n)$ is the VSS which is given by

$$\mu_{n_r}(n) = \mu_{\max}\cdot\frac{p_{n_r}^T(n)p_{n_r}(n)}{p_{n_r}^T(n)p_{n_r}(n)+C}, \quad (11)$$

where $C$ is a positive threshold parameter which is related to received signal-to-noise ratio (SNR), $C\sim\mathcal{O}(1/SNR)$. According to Eq. (11), the range of VSS is given as $\mu_{n_r}(n)\in(0,\mu_{\max})$, where $\mu_{\max}$ is the maximal step-size of gradient descend. Theoretically, the maximal step-size is less than 2 to ensure the adaptive algorithm stability [13]. Please note that $p_{n_r}(n)$ in Eq. (11) is given by

$$p_{n_r}(n) = \beta p_{n_r}(n-1) + (1-\beta)\frac{e_{n_r}(n)x(t)}{x^T(t)x(t)}, \quad (12)$$

where $\beta\in[1,0)$ is a smoothing factor to trade off VSS and estimation error.

*C. RZA-VSS-NLMS*

By introducing the VSS (11) into Eq. (9), improved sparse channel estimator $\tilde{h}_{n_r:}(n)$ is given by

$$\tilde{h}_{n_r:}(n+1) = \tilde{h}_{n_r:}(n) + \frac{\mu(n)e_{n_r}(n)x(t)}{x^T(t)x(t)} - \frac{\gamma_{RZA}\,\mathrm{sgn}\!\left(\tilde{h}_{n_r:}(n)\right)}{1+\varepsilon_{RZA}\left|\tilde{h}_{n_r:}(n)\right|}. \quad (13)$$

Please note that the second term in Eq. (13) attracts the channel coefficients $\tilde{h}_{n_r,n_t,l}(n)$, $l = 0,1,\cdots,L-1$ whose magnitudes are comparable to $1/\varrho_{RZA}$ to zeros. According the two proposed filtering algorithms in Eqs. (10) and (13), adaptive channel estimation methods for estimating MIMO channels are summarized in **Algorithm** 1.

---

**Input:** 1) $x(t)$ and $y(t)$;
        2) $\mu_{\max} = 1$ and $C$;
        3) $\lambda_{ZA}$ for ZA-VSS-NLMS
        4) $\lambda_{RZA}$ and $\varepsilon_{RZA}$ for RZA-VSS-NLMS.
**Output**: channel estimator $\tilde{H}$.
$n\leftarrow 1$; $p_{n_r}(0)\leftarrow 0$; $\tilde{H}(0)\leftarrow 0$;
**While** $\left\|\tilde{H}(n+1)-\tilde{H}(n)\right\|_2^2 \le 10^{-5}$ or $n\ge 5000$ **Do**
     $n_r\leftarrow \mathrm{mod}(n-1, N_r)+1$;
     $\tilde{h}_{n_r:}(n)\leftarrow \tilde{H}(n_r,:)$;
     $d_{n_r}(n)\leftarrow y(n_r)$;
     $e_{n_r}(n)\leftarrow d_{n_r}(n)-\tilde{h}_{n_r:}^T x(n)$;
     $p_{n_r}(n)\leftarrow \beta p_{n_r}(n-1)+(1-\beta)\dfrac{e_{n_r}(n)x(t)}{x^T(t)x(t)}$;
     $\mu_{n_r}(n)\leftarrow \mu_{\max}\cdot\dfrac{p_{n_r}^T(n)p_{n_r}(n)}{p_{n_r}^T(n)p_{n_r}(n)+C}$;
     $\tilde{h}_{n_r:}(n+1)\leftarrow \tilde{h}_{n_r:}(n)+\dfrac{\mu_{n_r}(n)e_{n_r}(n)x(t)}{x^T(t)x(t)}-\gamma_{ZA}\,\mathrm{sgn}(\tilde{h}_{n_r:}(n))$
     for ZA-VSS-NLMS in (10) or
     $\tilde{h}_{n_r:}(n+1)\leftarrow \tilde{h}_{n_r:}(n)+\dfrac{\mu(n)e_{n_r}(n)x(t)}{x^T(t)x(t)}-\dfrac{\gamma_{RZA}\,\mathrm{sgn}(\tilde{h}_{n_r:}(n))}{1+\varepsilon_{RZA}|\tilde{h}_{n_r:}(n)|}$
     For RZA-VSS-NLMS in (13);
**End**
$\tilde{H}(n_r,:)\leftarrow \tilde{h}_{n_r:}(n+1)$;

**Algorithm 1**: Sparse VSS-NLMS algorithms for estimating MIMO channels.

---

**Remark**: To better understanding the difference between ISS and VSS, based on Eqs. (8) and (10), it is worth mentioning that step size $\mu$ for sparse ISS-NLMS algorithm is invariable but the step size $\mu_{ZA}(n)$ for sparse VSS-NLMS algorithm is variable as depicted in as Fig. 3, where the maximal step size and ISS are set as $\mu_{max} \in \{0.5, 1.0\}$ and $\mu \in \{0.5, 1.0\}$, respectively. From the figure, one can easily find that ISS is kept invariant, while VSS $\mu(n)$ decreases as the estimation performance increases and vice versa. In other words, sparse VSS-NLMS algorithms adopting VSS for adaptive gradient descend, large step-size is adopted to speed up convergence rate for reducing computational complexity; small step-size is adopted to ensure algorithm stable in the case of high-accuracy estimator for further improving estimation performance.

## IV. COMPUTER SIMULATIONS

To confirm the effectiveness of the proposed methods, two metrics, i.e., MSE and BER, are adopted for performance evaluation. Channel estimators are evaluated by average MSE which is defined by

$$\text{Average MSE}\{\tilde{H}(n)\} = E\{\|H - \tilde{H}(n)\|_2^2\}, \quad (14)$$

and system performance is evaluated by the BER metric which adopts different data modulation schemes, such as phase shift keying (PSK) and quadrature amplitude modulation (QAM). The results are averaged over 200 independent Monte-Carlo runs. The length of each channel vector $h_{n_r n_t}$, $n_r = 1, 2, \cdots, N_r$ $n_t = 1, 2, \cdots, N_t$ is set as equal length with $L = 16$ and corresponding number of dominant taps is set to $T \in \{1, 4\}$. Each dominant channel tap follows random Gaussian distribution as $\mathcal{CN}(0, \sigma_h^2)$ and their positions are randomly decided within the length of $h_{n_r n_t}$. In addition, MISO channel vector $h_{n_r:}$ is subject to $E\|h_{n_r:}\|_2^2 = 1$. The received SNR is defined as $P_0/\sigma_n^2$, where $P_0$ is the power of received signal. Based on the research work in [14], it is worth mentioning that threshold parameters of sparse VSS-NLMS algorithm are adopted $C = 10^{-4}$ for 5dB and $C = 10^{-5}$ for 10dB and 20dB, respectively.

In the first example, average MSE performance of proposed methods is evaluated in Figs. 4~6 under two SNR regimes (i.e., 10dB and 20dB) in the case of $T = 1$ and 4, respectively. The effectiveness of the two proposed methods are confirmed when compared with previous methods, i.e., ISS-NLMS [13], VSS-NLMS [12] and sparse ISS-NLMS [5]–[8]. In addition, one can also find that two proposed methods depend channel sparsity as well as regularization parameter. Hence, to achieve a better steady-state estimation performance, regularization parameters for sparse VSS-NLMS algorithms, i.e., ZA-VSS-NLMS and RZA-VSS-NLMS, are adopted from the paper [15], which depends on the number of nonzero taps of a channel. In addition, since sparse VSS-NLMS algorithms take advantage of the channel sparsity as for prior information, hence, they achieve better estimation performance than standard VSS-NLMS algorithm, especially in a very sparse channel case, e.g., $T = 1$.

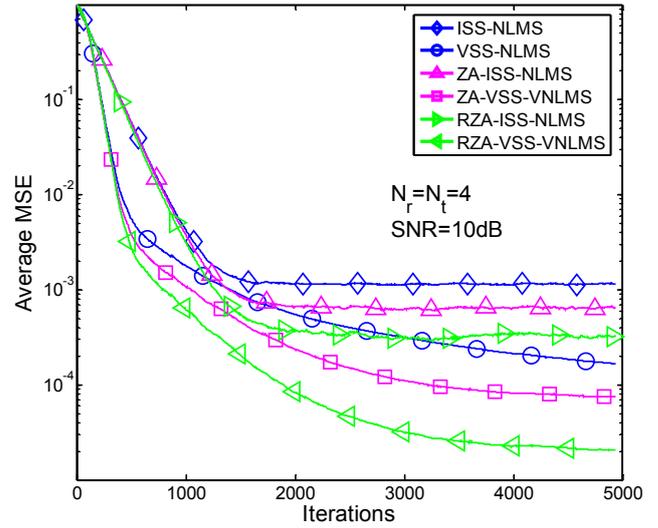

Fig. 4. Average MSE performance versus iterations ($T = 1$).

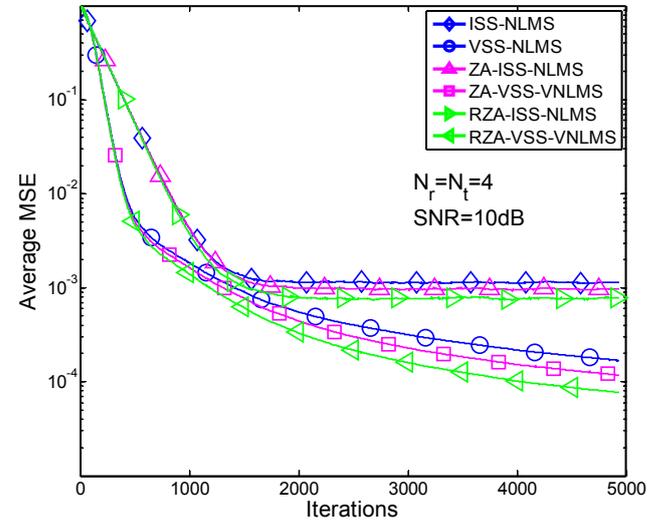

Fig. 5. Average MSE performance versus iterations ($T = 4$).

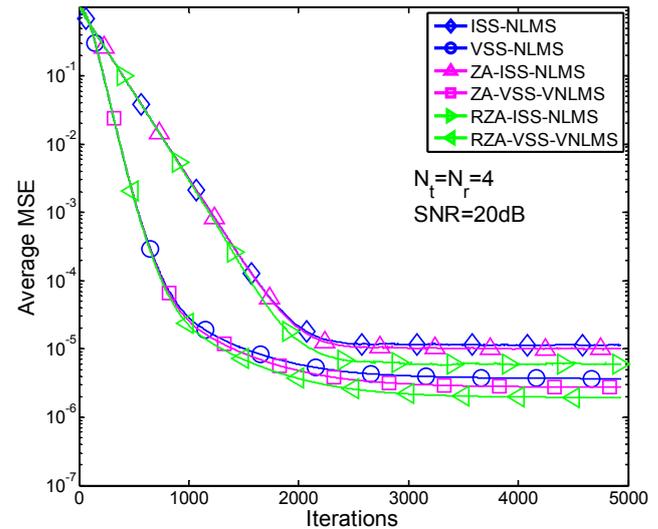

Fig. 6. Average MSE performance versus iterations ($T = 1$).

In the second example, system performance using proposed channel estimators is also evaluated with respect to BER performance. Multiple QAM schemes are considered. Received SNR is defined by $E_s/N_0$, where $E_s$ is the average received power of symbol and $N_0$ is the noise power. In Fig. 7, multiple QAM schemes, i.e., 16QAM, 64QAM and 128QAM, are considered for data modulation. One can easily find that the proposed method can achieve a better estimation than previous methods.

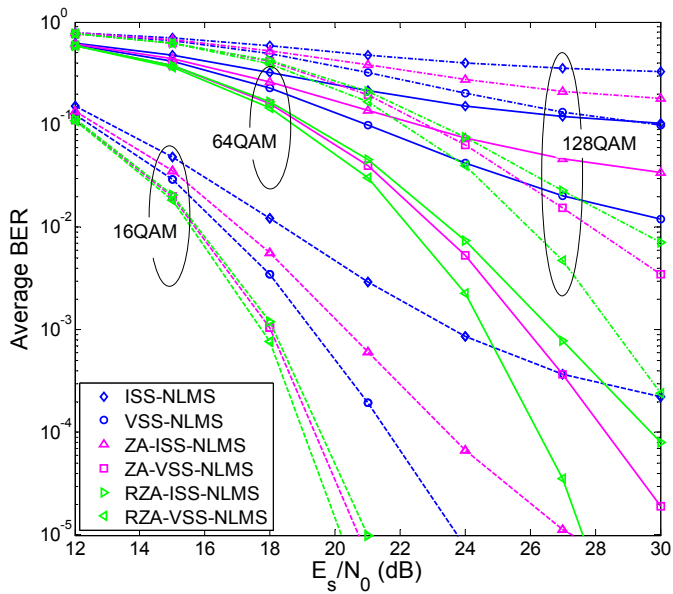

Fig. 7. Average BER performance versus received SNR (QAM).

## V. CONCLUSIONS AND FUTURE WORK

Traditional adaptive MIMO channel estimation methods tend to utilize the sparse ISS-NLMS algorithms. One of the main disadvantages of the traditional methods is the inability to balance the convergence speed and the estimation accuracy on the adaptive channel estimation. In this paper, two sparse VSS-NLMS algorithms, i.e., ZA-VSS-NLMS and RZA-VSS-NLMS, were proposed for estimating MIMO channels. Unlike the traditional sparse ISS-NLMS algorithms, the proposed algorithms utilized VSS which learns the estimation error and changes adaptively. Simulation results were provided to confirm the effectiveness of the proposed methods in three aspects convergence speed, estimation performance and system performance. First, convergence speed of sparse VSS-NLMS based methods is faster than ISS-NLMS based methods due to the fact that VSS for adaptive gradient descend is more efficient than ISS. Second, the proposed adaptive estimators can achieve better MSE gain than traditional methods in different SNR regimes especially for sparser channels. At last, system performance using the proposed channel estimators can also achieve better BER performance than previous methods especially in high-order modulation signal based MIMO communications systems.

Since the empirical parameter $C$ is adopted for the proposed sparse VSS-NLMS algorithm in Monte Carlo runs, it may cause the performance loss in different SNR regimes. In future work, we plan to develop an adaptive $C$ for the proposed algorithms so that it can learn the estimation error and SNR while without sacrificing much computation complexity.